\documentclass[showpacs,preprintnumbers,amsmath,amssymb]{revtex4}
\usepackage{graphicx}
\usepackage{dcolumn}
\usepackage{bm}

\begin{document}
\title{Calculation of the Superconductivity Gap of Metal from Its Parameters in Normal State}
\author{I. M. Yurin}
\affiliation{Institute of Physical Chemistry, Leninskiy prosp. 31,
GSP-1, Moscow, 119991, Russia}

\begin{abstract}
A previously considered model interpreted a superconductor as an
electron gas immersed in a medium with the dielectric constant
$\varepsilon$ and a certain elasticity, which could be determined by
measured sonic speed in the metal. The obtained expression of
effective electron-electron interaction (EEI) potential
unambiguously implied that, contrary to the suggestions of BCS
theory, it is its long-wave limit which is responsible for the
emergence of bound two-electron states and, consequently, for gap
formation in one-electron spectrum of the metal. However, the
existence of singularities in the EEI potential expression continued
to pose a problem, which did not allow a calculation of the gap
value for specific superconducting materials, first of all, for
metals belonging to periodic table (PT). In the present work, I
suggest taking into account matrix elements traditionally attributed
to electron scattering in EEI effective potential calculations. For
superconductors that has been made on the basis of a semiconductor
material by implanting electro-active defects into it, this
inclusion results in the appearance of an uncertainty of electron
momentum $\delta p \sim l^{ - 1}$, where $l$ is the electron free
path. When considering pure PT metals, Hamiltonian terms relating to
creation and annihilaton of phonons should be taken into account,
which also produces an uncertainty of electron momentum. This
uncertainty results in a regularization of EEI potential expression
and, therefore, in a possibility of examination of physical
properties of specific superconductors. Results of calculation of
superconductivity gap value for a range of simple metals (Al, Zn,
Pb, Sn) confirm the consistency of the developed approach.
\end{abstract}
 \maketitle
\section{Introduction}
According to the opinion expressed by the most part of the
scientific community, the present-day situation of the
superconductivity theory can hardly be called satisfactory. Indeed,
while the BCS theory \cite{1} provides a qualitative explanation of
phenomena observed in superconductors, it is unable to produce
reliable predictions of their quantitative parameters. At the same
time, it is implicitly suggested that the BCS theory contains no
contradictions in its description of the 'conventional'
superconductivity; problems are only believed to appear in
connection with the description of high-temperature
superconductivity.

That is why recent studies mostly concern high-temperature
superconductors alone \cite{2,3}. All these studies include the BCS
theory considered as an extreme case; they only differ in mechanisms
producing the attraction between electrons. Therefore, these models
do not essentially change the microscopic picture of the phenomenon.

According to the concept of the so called BCS-BEC crossover
\cite{4}, the phase transition in the BCS theory shares many
features with the Bose-Einstein condensation (BEC) phenomenon,
which, eventually, allowed connecting it with the Ginzburg-Landau
theory \cite{5}. Currently, attempts are being made at providing a
more precise description of superconductors within the framework of
this concept. These models also incorporate the BCS model for an
extreme case.

It should be noted that, in 1980-90s, superconductivity theories
unrelated to the BCS theory have been put forward. Most of these
models contradict experimental data even at the level of the
qualitative description. By the present date, the interest towards
such models has seen a substantial decrease.

The superconductor model to be described in the present paper does
not contain the BCS model as an extreme case either, and has nothing
in common with the BEC phenomenon. Nevertheless, the model under
development can already claim, at least, a semi-quantitative
agreement with the experiment. Its present-day status is described
in Sections 2 and 3. This approach has been discussed in detail in a
number of publications \cite{6,7}; therefore, here we restrict
ourselves to a shortest possible description, only indicating its
initial differences with currently universally acknowledged
concepts.

The goal of the present study is a calculation of the
superconducting gap in the framework of the proposed model in order
to compare it with the analogous calculation results of the BCS
theory. The study is performed on simple metals belonging to PT.

Fitting parameters obtained out of the quantitative calculation have
the order of magnitude corresponding to that predicted by the
theory. This agreement makes the approach under consideration look
more attractive than the BCS theory.

\section{Renormalization procedure for metals}
A procedure of renormalization for a system comprised of electrons
and phonons correctly taking into account the singular nature of
matrix elements of electron-phonon interaction (EPI) in the
long-wave limit has been performed previously \cite{6}. The
Hamiltonian of a monatomic metal electron system, as obtained from
the renormalization, has the following form:
\begin{equation}
H = H_0  + H_{ee}
\end{equation}
where
\begin{equation}
H_0  = \sum\limits_{\sigma ,{\bf p}} {t_{\bf p} c_{\sigma ,{\bf p}}^
+  c_{\sigma ,{\bf p}} } ,
\end{equation}
\begin{equation}
H_{ee}  = \Omega ^{ - 1} \sum\limits_{\sigma ,{\bf p}}
{\sum\limits_{\nu ,{\bf k}} {\sum\limits_{\bf q} {U_{\bf q}^{{\bf
p},{\bf k}} c_{\sigma ,{\bf p} + {\bf q}}^ +  c_{\nu ,{\bf k} - {\bf
q}}^ +  c_{\nu ,{\bf k}} c_{\sigma ,{\bf p}} } } } ,
\end{equation}
\begin{equation}
U_{\bf q}^{{\bf p},{\bf k}}  = \frac{{2\pi e^2 }}{{\varepsilon q^2
}} + \delta U_{\bf q}^{{\bf p},{\bf k}} ,
\end{equation}
\begin{equation}
\delta U_{\bf q}^{{\bf p},{\bf k}}  = \frac{{\pi ze^2 K_F^2
}}{{3\varepsilon mM}}\left[ {\frac{1}{{\left( {t_{{\bf p} + {\bf q}}
- t_{\bf p} } \right)^2  - S_1^2 q^2 }} + \frac{1}{{\left( {t_{\bf
k}  - t_{{\bf k} - {\bf q}} } \right)^2  - S_1^2 q^2 }}} \right],
\end{equation}
$c^+$ and $c$ are creation and annihilation operators, respectively,
of conductivity band electrons; $\sigma$ and $\nu$ electron spin
indexes; $ z$ is the number of conductivity electrons per atom; $m$
is the electron mass; $t_{\bf p}  = \frac{{p^2 }}{{2m}}$; $M$ is the
ion mass; $S_1$ is a parameter related to the observed sonic speed
of the longitudinal polarization $S$ as $S_1^2 = S^2  -
\frac{{zK_F^2 }}{{6mM}}$; $K_F$ is Fermi vector; $\Omega = L \times
L \times L$; $L$ is the crystal dimension, and $\varepsilon$ is the
statical long-wave limit of the metal dielectric function caused by
interband transitions.

The above expression for the initial potential of EEI substantially
differs from other known results, first of all because $\delta U
\sim q^{ - 2}$ at $q \to 0$. Evidently, this result cannot be
obtained within the framework of Fr\"ohlich model \cite{8} as this
model lacks a consistent accounting for screening effects when
calculating the EPI Hamiltonian terms. This circumstance has long
been known \cite{9}; now it is worthwhile to note here that it is
the Fr\"ohlich model that makes a basis for Eliashberg equations,
which the majority of authors \cite{10} consider to represent the
most integral superconductor description in BCS theory.

The expression (4) obtained  for the initial EEI excludes the
formation of bound two-electron states in the system at this stage
of problem examination. Indeed, the Coulomb term is significantly
larger than $\delta U$, since $\delta U \to 0$ at $M \to \infty$. As
a formation of bound two-electron states of the Hamiltonian (1)
appears to be impossible, any 'bosonization' of the system is
completely out of question. That is why the models involving BEC to
clarify the superconductivity phenomenon \cite{11,12} seem to be
insufficiently substantiated.

\section{Excitation spectra in multi-particle systems}

Let us assume by definition that the Hamiltonian $\tilde H$,defined
as:
\begin{eqnarray}
 \tilde H &=& \sum\limits_{\sigma ,{\bf p}} {t_p c_{\sigma ,{\bf p}}^ +  c_{\sigma ,{\bf p}} }
  + \Omega ^{ - 1} \sum\limits_{\sigma ,{\bf p}} {\sum\limits_{\nu ,{\bf k}} {\sum\limits_{\bf q} {\tilde U_{\bf q}^{{\bf p},{\bf k}} c_{\sigma ,{\bf p} + {\bf q}}^ +  c_{\nu ,{\bf k} - {\bf q}}^ +  c_{\nu ,{\bf k}} c_{\sigma ,{\bf p}} } } }
 \end{eqnarray}
has the same two-electron excitation spectrum as the initial
Hamiltonian (1) accounting for multi-particle effects. Thus, in
fact, we rigorously define the notion of the EEI effective potential
$\tilde U_{\bf q}^{{\bf p},{\bf k}}$ involved in the right-hand side
of Eq. (6).

We start the procedure of definition of $\tilde U_{\bf q}^{{\bf
p},{\bf k}}$ by considering the following transformation:
\begin{equation}
\begin{array}{l}
 C_{\sigma ,{\bf p}}^ +   = c_{\sigma ,{\bf p}}^ +   + \sum\limits_\nu  {\sum\limits_{{\bf k},{\bf q}} {\theta _{\sigma ,\nu ,{\bf q}}^{{\bf p},{\bf k}} c_{\sigma ,{\bf p} + {\bf q}}^ +  c_{\nu ,{\bf k} - {\bf q}}^ +  c_{\nu ,{\bf k}} } } , \\
 C_{\sigma ,{\bf p}}  = c_{\sigma ,{\bf p}}  + \sum\limits_\nu  {\sum\limits_{{\bf k},{\bf q}} {\theta _{\sigma ,\nu ,{\bf q}}^{*{\bf p},{\bf k}} c_{\nu ,{\bf k}}^ +  c_{\nu ,{\bf k} - {\bf q}} c_{\sigma ,{\bf p} + {\bf q}} } } ,
 \end{array}
\end{equation}
where
\begin{equation}
\begin{array}{l}
 \theta _{\sigma ,\sigma ,{\bf q}}^{{\bf p},{\bf k}}  = \frac{1}{2}\left( {\delta _{{\bf k} - {\bf p}}^{\bf q}  - \delta _{\bf q}^0 } \right) + \chi _{1,{\bf q}}^{{\bf p},{\bf k}} , \\
 \theta _{\sigma , - \sigma ,{\bf q}}^{{\bf p},{\bf k}}  =  - \delta _{\bf q}^0  + \chi _{0,{\bf q}}^{{\bf p},{\bf k}}  + \chi _{1,{\bf q}}^{{\bf p},{\bf k}} ,
 \end{array}
\end{equation}
$\chi _{0,{\bf q}}^{{\bf p},{\bf k}}$ and $\chi _{1,{\bf q}}^{{\bf
p},{\bf k}}$ are expressed in terms of wave functions (WFs) ) of
stationary two-particle states $\left| {S_0^{{\bf p},{\bf k}} }
\right\rangle  = \sum\limits_{\bf q} {\chi _{0,{\bf q}}^{{\bf
p},{\bf k}} \left( {c_{ \uparrow ,{\bf p} + {\bf q}}^ +  c_{
\downarrow ,{\bf k} - {\bf q}}^ +   - c_{ \downarrow ,{\bf p} + {\bf
q}}^ +  c_{ \uparrow ,{\bf k} - {\bf q}}^ +  } \right)}$ and $\left|
{S_1^{{\bf p},{\bf k}} } \right\rangle  = \sum\limits_{\bf q} {\chi
_{1,{\bf q}}^{{\bf p},{\bf k}} c_{ \uparrow ,{\bf p} + {\bf q}}^ +
c_{ \uparrow ,{\bf k} - {\bf q}}^ +  } \left| 0 \right\rangle$ in
the following way:
\begin{equation}
\begin{array}{l}
 \tilde H\left| {S_1^{{\bf p},{\bf k}} } \right\rangle  = \left( {t_{\bf p}  + t_{\bf k}  + \delta E_1^{{\bf p},{\bf k}} } \right)\left| {S_1^{{\bf p},{\bf k}} } \right\rangle , \\
 \tilde H\left| {S_0^{{\bf p},{\bf k}} } \right\rangle  = \left( {t_{\bf p}  + t_{\bf k}  + \delta E_0^{{\bf p},{\bf k}} } \right)\left| {S_0^{{\bf p},{\bf k}} } \right\rangle .
 \end{array}
\end{equation}
It is easy to demonstrate that the transformation (7) reduces the
Hamiltonian $\tilde H$ to the diagonal form in the two-particle
state space:
\begin{eqnarray}
 \tilde {H} &=& \sum\limits_{\sigma ,{\bf k}} {t_{\bf k} C_{\sigma ,{\bf k}}^ +  C_{\sigma ,{\bf k}} }\nonumber\\& &
  + \frac{1}{2}\sum\limits_{\sigma ,{\bf p} \ne {\bf k}} {\delta E_1^{{\bf p},{\bf k}} C_{\sigma ,{\bf p}}^ +  C_{\sigma ,{\bf k}}^ +  C_{\sigma ,{\bf k}} C_{\sigma ,{\bf p}} }  + \frac{1}{4}\sum\limits_{{\bf p} \ne {\bf k}} {\delta E_1^{{\bf p},{\bf k}} \left( {C_{ \uparrow ,{\bf p}}^ +  C_{ \downarrow ,{\bf k}}^ +   + C_{ \downarrow ,{\bf p}}^ +  C_{ \uparrow ,{\bf k}}^ +  } \right)\left( {C_{ \downarrow ,{\bf k}} C_{ \uparrow ,{\bf p}}  + C_{ \uparrow ,{\bf k}} C_{ \downarrow ,{\bf p}} } \right)}  \nonumber\\& &
  + \sum\limits_{\bf p} {\delta E_0^{{\bf p},{\bf p}} C_{ \uparrow ,{\bf p}}^ +  C_{ \downarrow ,{\bf p}}^ +  C_{ \downarrow ,{\bf p}} C_{ \uparrow ,{\bf p}} }  + \frac{1}{4}\sum\limits_{{\bf p} \ne {\bf k}} {\delta E_0^{{\bf p},{\bf k}} \left( {C_{ \uparrow ,{\bf p}}^ +  C_{ \downarrow ,{\bf k}}^ +   - C_{ \downarrow ,{\bf p}}^ +  C_{ \uparrow ,{\bf k}}^ +  } \right)\left( {C_{ \downarrow ,{\bf k}} C_{ \uparrow ,{\bf p}}  - C_{ \uparrow ,{\bf k}} C_{ \downarrow ,{\bf p}} } \right)} .
\end{eqnarray}
Taking into account multi-particle effects in electron gas is
performed in the random phase approximation (RPA) as follows.
Consider the inverse transformation of (7):
\begin{equation}
\begin{array}{l}
 c_{\sigma ,{\bf p}}^ + = C_{\sigma ,{\bf p}}^ +   + \sum\limits_\nu  {\sum\limits_{{\bf k},{\bf q}} {\theta _{\sigma ,\nu ,{\bf q}}^{*{\bf k} - {\bf q},{\bf p} + {\bf q}} C_{\sigma ,{\bf p} + {\bf q}}^ +  C_{\nu ,{\bf k} - {\bf q}}^ +  C_{\nu ,{\bf k}} } } , \\
 c_{\sigma ,{\bf p}}= C_{\sigma ,{\bf p}}  + \sum\limits_\nu  {\sum\limits_{{\bf k},{\bf q}} {\theta _{\sigma ,\nu ,{\bf q}}^{{\bf k} - {\bf q},{\bf p} + {\bf q}} C_{\nu ,{\bf k}}^ +  C_{\nu ,{\bf k} - {\bf q}} C_{\sigma ,{\bf p} + {\bf q}} } } .
 \end{array}
\end{equation}
Then express the initial Hamiltonian (1) in terms of operators $C^
+$ and $C$ using the transformations (11). In this form, the
Hamiltonian comprises terms of the structure $C^ +  C^ +  C^ + CCC$,
which, in the RPA, can be reduced to the form of $\left\langle {C^ +
C} \right\rangle C^ + C^ +  CC$. These terms define additional
potentials $\tilde D_{\bf q}^{{\bf p},{\bf k}}$, which can be
included in the equation for effective potentials $ \tilde U_{\bf
q}^{{\bf p},{\bf k}}$:
\begin{equation}
\tilde U_{\bf q}^{{\bf p},{\bf k}}  = U_{\bf q}^{{\bf p},{\bf k}}  +
\tilde D_{\bf q}^{{\bf p},{\bf k}} .
\end{equation}
Eqs. (9) and (12) form a self-consistent system  for $\tilde U_{\bf
q}^{{\bf p},{\bf k}}$ definition.

\emph{It is important to note that, in the framework of the proposed
approach, it is the Hamiltonian (10) which is related to the initial
Hamiltonian (1) by the unitary transformation (7), and the
introduction of the Hamiltonian (6) into the theory is of an
auxiliary nature and is used to calculate the parameters $\delta
E_i^{{\bf p},{\bf k}}$ and  $\theta _{\sigma ,\nu ,{\bf q}}^{{\bf
p},{\bf k}}$.}

We should specify more explicitly in which sense we name the
transformation (7) unitary, since the anti-commutators $ \left[
{C^+,C}\right]_ +$ and $ \left[ {C,C} \right]_+$ differ from
standard fermion commutation relations in the presence of terms
$c^+c^+cc$ and $c^+ccc$, respectively \cite{7}. It is, however, easy
to demonstrate that, in two-electron state space, a transition from
$c_{\sigma ,{\bf p}}^ + c_{\nu ,{\bf k}}^ + \left| 0 \right\rangle$
states to $C_{\sigma ,{\bf p}}^ +  C_{\nu ,{\bf k}}^ +  \left| 0
\right\rangle$ does not alter the WFs scalar products; therefore,
the transformation (7) for this case should be considered to be
unitary.

Then, the right-hand sides of Eqs. (6,7,11) can be considered as two
first terms of the series with the number of terms equal to the
number of particles in the system. These series are composed while
determining the multi-particle excitation spectrum using the same
procedure as invoked above for the calculation of a two-electron
excitation spectrum. Of course, the transformation (7) completed
with the terms of the corresponding series is unitary for all
states, in which the number of particles is less than or equal to
the number of particles in the system.

Using the RPA for the examination of an excitation spectrum allows
for an introduction the Hartree-Fock method for calculating of
one-electron energies into the above calculation routine. From this
point of view, the proposed approach is essentially a generalization
of the Hartree-Fock approximation for a multi-particle spectrum
calculation.

Previously \cite{6}, the above mentioned self-consistent system for
the potential (4) have been formulated and solved using the
perturbation theory for solution of eigenvalue problems (9). It
should also be noted that considering the Coulomb EEI alone (see
Appendix A) within the framework of the proposed scheme is well
consistent with the modern idea of the effective EEI in a degenerate
plasma. This fact confirms the consistency of the proposed approach
for treatment of multi-particle problems with interaction.

Combining (A5) with previously obtained results \cite{6} yields the
following equation:
\begin{eqnarray}
 \tilde U_{\bf q}^{{\bf p},{\bf k}}  &=& 2\pi e^2 \frac{{\varepsilon q^2 }}{{\left( {\varepsilon q^2  + \lambda ^2 } \right)^2 }}
  - 4\left( {\frac{{zm}}{{3M}}} \right)^2 \frac{{2\pi e^2 }}{{\varepsilon q^2 }}\frac{{K_F^4 }}{{\left( {q^2  - \chi } \right)^2 }},
\end{eqnarray}
where $\chi  = 4m^2 \left( {S^2  - \frac{{zK_F^2 }}{{3mM}}}
\right)$. In this form, the equation for $\tilde U_{\bf q}^{{\bf
p},{\bf k}}$ is valid in two limits: $M \to \infty$ and $q \to 0$.

It ensues from Eq. (13) that an attraction between electrons is
present in the long-wave limit. Now according to Eliashberg
equations, Cooper pair formation is mainly due to short-wave phonons
with a wave vector $q \approx 2K_F$. Evidently, the authors of the
theory were guided by the erroneous hypothesis of absence of
electrons attraction in the long-wave limit, which had been
suggested based on the plasma model analysis performed back in
1950's \cite{13} .

If Eqs. (9) do not allow the appearance of bound states, then
$\delta E_i^{{\bf p},{\bf k}}  \sim \Omega ^{ - 1}$, and the theory
describes a normal metal. Appearance of bound states with the
binding energy $E_b$ allows to simplify the Hamiltonian (10).
Neglecting energy corrections $\delta E_i^{{\bf p},{\bf k}} \sim
\Omega ^{ - 1}$ for unbound states of electron pairs, the
Hamiltonian (10) for a superconductor can be written as \cite{7}:
\begin{eqnarray}
 H &=& \sum\limits_{\sigma ,{\bf k}} {t_{\bf k} C_{\sigma ,{\bf k}}^ +  C_{\sigma ,{\bf k}} }  - \sum\limits_{\bf k} {E_b C_{ \uparrow ,{\bf k}}^ +  C_{ \downarrow ,{\bf k}}^ +  C_{ \downarrow ,{\bf k}} C_{ \uparrow ,{\bf k}} }
 \nonumber\\& &
  - \frac{1}{4}\sum {E_b \left( {C_{ \uparrow ,{\bf p}}^ +  C_{ \downarrow ,{\bf k}}^ +   - C_{ \downarrow ,{\bf p}}^ +  C_{ \uparrow ,{\bf k}}^ +  } \right)\left( {C_{ \downarrow ,{\bf k}} C_{ \uparrow ,{\bf p}}  - C_{ \uparrow ,{\bf k}} C_{ \downarrow ,{\bf p}} } \right)}
  \nonumber\\& &
\left\{ {{\bf p} \ne {\bf k};\left| {{\bf p}_\alpha   - {\bf
k}_\alpha  } \right| \le 2\pi /L,\alpha  = x,y,z} \right\}
\end{eqnarray}

It is easily seen that the ground state $\left| {\Phi _0 }
\right\rangle$ of the Hamiltonian (14) coincides with the well-known
expression for a non-interacting Fermi gas:
\begin{equation}
\left| {\Phi _0 } \right\rangle  = \prod\limits_{p < K_F } {C_{
\uparrow ,{\bf p}}^ +  C_{ \downarrow ,{\bf p}}^ +  } \left| 0
\right\rangle .
\end{equation}
Therefore, from the point of view of the new theory, there is no
ground for suggesting a possibility of formation of BEC, although
the theory does discuss wave functions of bound states of electron
pairs. Hence, the Ginsburg-Landau theory \cite{5} also lacks
microscopic grounds. Nevertheless, we take it as evident that main
predictions of the Ginsburg-Landau theory are reproduced within the
framework of the proposed theory; therefore, the absence of BEC in
the system in question is not to be worried about.

Along with the ground state $\left| {\Phi _0 } \right\rangle$, also
the one-electron and one-hole excitations are the stationary states
of Hamiltonian (14). This allows to calculate the arising energy gap
value $2\Delta _0$ in the one-electron spectrum of the ground state.
If there exists only one bound solution with a zero spin and a fixed
pair momentum, the value of $\Delta _0$   is calculated as:
\begin{equation}
\Delta _0  = 7E_b .
\end{equation}

\section{Regularization of EEI effective potential}
The expression (13) is evidently inapplicable for calculation of
superconductive properties of metals due to the presence of
singularities at $q = 0$ and $q = \sqrt \chi$. On the other hand, we
have not up to now taken into account the Hamiltonian terms usually
related to scattering processes, which can result in an uncertainty
of electron momentum and, consequently, in a removal of
singularities even in the initial EEI potential $U_{\bf q}^{{\bf
p},{\bf k}}$.

Let us begin with the simplest case of a metal that has been made on
the basis of a semiconductor material by implanting electro-active
defects into it. Evidently, in this case, the abovementioned
momentum uncertainty is $\delta p \sim l^{ - 1}$, where $l$ is the
free path of electrons as determined by the scattering on lattice
defects.

In the case of pure PT metals, the following considerations should
be taken into account. The renormalization procedure for metals
\cite{6} used only longitudinal phonons with a distinctly expressed
singularity of the EPI potential in the long-wave limit. However,
due to various reasons, the EPI terms remaining in the Hamiltonian
after this renormalization have a weak non-analyticity in the said
limit for acoustic phonons. Neglecting relativistic effects these
terms have the following form:
\begin{equation}
H_{ep}  = \sum\limits_{\sigma ,{\bf p}} {\sum\limits_{m,{\bf
q},{\bf g}} {\sqrt {\omega _{m,{\bf q}} } F_{m,{\bf q}}^{{\bf
p},{\bf g}} c_{\sigma ,{\bf p} - {\bf q} - {\bf g}}^ +  c_{\sigma
,{\bf p}} b_{m,{\bf q}}^ +  } }  + H.c.,
\end{equation}
where $\omega _{m,{\bf q}}$ is the phonon frequency; $b_{m,{\bf q}}^
+$ and $b_{m,{\bf q}}$ are phonon creation and annihilation
operators, respectively, including the case of transversal
polarization, $m$ and ${\bf q}$ are the phonon mode index and phonon
quasi-momentum, respectively, and ${\bf g}$ is the reciprocal
lattice vector. As to  $F_{m,{\bf q}}^{{\bf p},{\bf g}}$, it is a
piecewise regular function with respect to variations of the
arguments  ${\bf p}$ and ${\bf q}$.

Taking into account the term $H_{ep}$ radically alters the WFs of
stationary one-electron states due to the possibility of spontaneous
creation of phonons from electrons, since Fermi speed is much
greater than the sonic speed. Therefore, from the point of view of
calculation of the effective EEI, the phonon system acts as a
reservoir that electrons can exchange momentum with, which can be
accounted for by a regularization of EEI potential expressions.

Let us try to determine the arising momentum uncertainty from the
consideration of the temperature dependence of the specific
resistance  $\rho$ at temperatures $T \gg \theta _D$, where $\theta
_D$ is Debye temperature. Then, according to the Fermi's Golden
Rule, the following relation is valid for electron-phonon transition
frequency $\nu _T$ of an electron of momentum  ${\bf p}$:
\begin{equation}
\nu _T  \sim T\sum\limits_{m,{\bf g}} {\int {\left( {{\bf q} + {\bf
g}} \right)^2 \left| {F_{m,{\bf q}}^{{\bf p},{\bf g}} } \right|^2
\delta \left( {t_{\bf p}  - t_{{\bf p} - {\bf q} - {\bf g}}  -
\omega _{m,{\bf q}} } \right)d{\bf q}} } ,
\end{equation}
which takes into account that in this case, the transitions are
induced by phonons with a distribution function approximately equal
to $T/\omega _{m,{\bf q}}$.

Of course, what we are interested in is the spontaneous transition
frequency $\nu _0$. It can be expressed as:
\begin{equation}
\nu _0  \sim \sum\limits_{m,{\bf g}} {\int {\left( {{\bf q} + {\bf
g}} \right)^2 \left| {\sqrt {\omega _{m,{\bf q}} } F_{m,{\bf
q}}^{{\bf p},{\bf g}} } \right|^2 \delta \left( {t_{\bf p}  -
t_{{\bf p} - {\bf q} - {\bf g}}  - \omega _{m,{\bf q}} }
\right)d{\bf q}} } .
\end{equation}
Comparison of  Eqs. (18) and (19) yields:
\begin{equation}
\frac{{\nu _0 }}{{\nu _T /T}} = \bar \omega ,
\end{equation}
where
\begin{equation}
\bar \omega  = {{\sum\limits_{m,{\bf g}} {\int {\left( {{\bf q} +
{\bf g}} \right)^2 \omega _{m,{\bf q}} \left| {F_{m,{\bf q}}^{{\bf
p},{\bf g}} } \right|^2 \delta \left( {t_{\bf p}  - t_{{\bf p} -
{\bf q} - {\bf g}}  - \omega _{m,{\bf q}} } \right)d{\bf q}} } }
\over {\sum\limits_{m,{\bf g}} {\int {\left( {{\bf q} + {\bf g}}
\right)^2 \left| {F_{m,{\bf q}}^{{\bf p},{\bf g}} } \right|^2 \delta
\left( {t_{\bf p}  - t_{{\bf p} - {\bf q} - {\bf g}}  - \omega
_{m,{\bf q}} } \right)d{\bf q}} } }}.
\end{equation}
It is important to note that the following relationships are
strictly fulfilled for  $\bar \omega$:
\begin{equation}
0 < \bar \omega  < \omega _{\max } ,
\end{equation}
where $\omega _{\max }$ is the maximum frequency of the phonon
spectrum.

On the other hand, the Hamiltonian (17) is in certain cases the main
cause of energy dissipation in electron tunneling between
superconductors. That is why a correct numerical processing of the
current-voltage characteristics (CVC) in the region where the tunnel
junction (TJ) behaves itself similarly to resistance yields a more
precise evaluation of the parameter $\bar \omega$ than the
relationship (22).

The $\nu _T /T$ value can be roughly evaluated from Drude formula
for specific resistance $\rho$ at high temperatures:
\begin{equation}
\nu _T /T = \frac{{\Omega _{pl}^2 \rho '}}{{4\pi }}
\end{equation}
where $\Omega _{pl}$ is plasma frequency determined in optical
studies of metals in the infrared (IR) range, $ \rho '$ is the
derivative of the specific electric resistance over temperature, for
which, evidently, $\rho ' \approx \rho /T$. Considering that the
free path of a single electron in the system is calculated as  $l =
\left\langle {V_F } \right\rangle /\nu _0$, where $\left\langle {V_F
} \right\rangle$ is an average electron velocity over the Fermi
surface, the definitive expression for the sought uncertainty
$\delta p$ of a stationary single-electron state is:
\begin{equation}
\delta p = \frac{{\Omega _{pl}^2 \rho '\bar \omega }}{{4\pi
\left\langle {V_F } \right\rangle }}
\end{equation}

In the further study, taking into account that for the metals
considered in the present paper  $\chi  < 0$, we use the following
regularized expression of the EEI effective potential (cf. Eq.
(13)):
\begin{equation}
 \tilde U_{\bf q}^{{\bf p},{\bf k}}  = 2\pi e^2 \frac{{\varepsilon q^2 }}{{\left( {\varepsilon q^2  + \lambda ^2 } \right)^2 }}
  - 4\left( {\frac{{zm}}{{3M}}} \right)^2 \frac{{2\pi e^2 }}{{\varepsilon \left( {q^2  + \delta p^2 } \right)}}\frac{{K_F^4 }}{{\left( {q^2  + \sqrt {\chi ^2  + \delta p^4 } } \right)^2
  }}.
 \end{equation}
This version of regularization allows to obtain an analytical
expression for the EEI potential in a coordinate representation,
which simplifies the calculation procedure.

\section{Superconductivity gap calculations for PT metals}

Equation (25) is suitable for calculating the superconductivity gap.
From the technical point of view, the task can be reduced to
determining the energy of bound states of pairs for a potential that
is spherically symmetrical in the coordinate representation, whose
Fourier transform was defined in Eq.(25).

Considering the validity of the relationship $\lambda \gg \delta p$,
it can easily be concluded that in the coordinate representation the
first term of the right-hand side of Eq. (25), which describes a
screened Coulomb repulsion of electrons, has a form of a
delta-shaped function as compared with the second term. Therefore,
its existence can be accounted for virtually without affecting the
precision of the calculation by introducing the boundary condition
of $\psi \left( r \right) = 0$ at $r = 2\sqrt \varepsilon \lambda ^{
- 1}$ into the appearing eigenvalue problem:
\begin{equation}
 - \frac{{\Delta \psi }}{{2m}} + V\left( r \right)\psi \left( {\bf r} \right) =  - \frac{{E_b }}{2}\psi \left( {\bf r} \right),
\end{equation}
where
\begin{equation}
V\left( r \right) = \frac{1}{{2\pi ^2 }}\int {\frac{{\sin \left(
{qr} \right)}}{r}qU\left( q \right)dq} ,
\end{equation}
and $U\left( q \right)$ is equal to the second term of the
right-hand side of Eq. (25).

Having solved Eq. (26) we evaluated the appearing superconductivity
gap using Eq. (16) or its generalized form $\Delta _0  =
7\sum\limits_j {E_j }$, where $j$ is the number of the bound state
if Eq. (26) had multiple bound solutions, as in the case of Pb,
where two bound states are present, and the ratio of their binding
energies is approximately equal to 0.04.

Let us dwell on the problem of estimation of value $\left\langle
{V_F } \right\rangle$, introduced in Eq. (24). To estimate it, we
use the method similar to that suggested by G. P. Motulevich
\cite{14}. Namely, consider the parameter $\gamma _{opt}$, which
relates the plasma frequency value $ \Omega _{pl}$, as observed in
IR-range measurement to the corresponding value $\Omega _0$
calculated based on the metal valence:
\begin{equation}
\gamma _{opt}  = z\Omega _{pl}^2 /\Omega _0^2 .
\end{equation}
Assume that the approximate equation $\int {{\bf V}d{\bf S}_F }
\approx \left\langle {V_F } \right\rangle S_F$, where $S_F$ is the
area of the Fermi surface, is valid for the integral  $\int {{\bf
V}d{\bf S}_F }$ over the Fermi surface, where ${\bf V}$  is the
electron velocity. Then, according to Drude formula:
\begin{equation}
\left\langle {V_F } \right\rangle S_F  = \frac{{\gamma _{opt}
}}{z}V_F^0 S_F^0 ,
\end{equation}
where $V_F^0$ and $S_F^0$ are Fermi velocity and Fermi surface area
as calculated based on the metal valence, similarly to  $\Omega _0$.
On the other hand, considering the ratio of density of states at
Fermi level over its value calculated for an ideal electron gas of
the same concentration, we obtain:
\begin{equation}
S_F /\left\langle {V_F } \right\rangle  = \gamma _{el} S_F^0
/V_F^0 ,
\end{equation}
where the parameter $\gamma _{el}$ can be determined based on the
metal electronic specific heat data at low temperatures. Comparing
Eqs. (29-30) yields:
\begin{equation}
\left\langle {V_F } \right\rangle  = \sqrt {\frac{{\gamma _{opt}
}}{{z\gamma _{el} }}} V_F^0 ,
\end{equation}
which provides a solution of the problem as stated.

The only ill-defined parameter of the calculation is the value of
$\bar \omega$, which is only limited by the relationships (22). That
is why the proposed version of the calculation used this value as
fitting parameter. In other terms, the parameter $\bar \omega$ used
in the definition of $ \delta p$ was chosen such as to make the
resulting width of the superconductivity gap coincide with its
experimentally known  values.

Table ~\ref{tab:metals} lists both initial parameters ($z$,$ \gamma
_{opt}$, $\gamma _{el}$, $\varepsilon$, $\Delta _0$) and the
calculated values of the parameter $\bar \omega$ for Al, Zn, Pb, and
Sn. An analysis of the ratio $\bar \omega /\theta _D$, as provided
in the same table indicates that fitting parameters $\bar \omega$
have the same order of magnitude as the Debye temperature $\theta
_D$ in full accordance with Eq.(21). It should be noted here that
the initial data of the calculations had a fairly wide variation
range: $\left( {M^{\left\{{Pb}\right\}}/M^{\left\{{Al}
\right\}}}\right)^2\approx 60$, $\rho '^{\left\{{Pb}\right\}}/\rho
'^{\left\{{Al}\right\}}\approx 7$,
$\Delta_0^{\left\{{Pb}\right\}}/\Delta _0^{\left\{{Zn}
\right\}}\approx 9$, $\varepsilon^{\left\{ {Sn}\right\}}/\varepsilon
^{\left\{{Pb}\right\}}\approx 4.5$,
$\theta_D^{\left\{{Al}\right\}}/\theta_D^{\left\{{Pb}\right\}}\approx
4$. In view of this fact, a physicist would normally consider
accidental coincidences out of question, and acknowledge the
proposed theory.

This circumstance suggests that a discussion on the approximations
introduced within the framework of the approach under development,
involving experts in both theoretical physics and mathematics, would
be quite useful.

Figs. 1-4 illustrate the calculation results for the considered
metals.

\begin{table*}
\caption{\label{tab:metals}The initial and calculated parameters of
some simple PT metals. The values of $\rho '$ are given for the
temperature $T=291$ K.}
\begin{ruledtabular}
\begin{tabular}{ccccccccccc}
 Metal&z&$10^9 \rho '$, ${\rm Ohm} \cdot {\rm cm/K}$
&$10^{ - 5} S$, ${\rm cm/sec}$&$\gamma _{opt}$ \cite{14}&$\gamma
_{el}$&$\varepsilon$
 \cite{15,16,17}&$\Delta _0$, K&$\bar \omega$, K&$\theta _D$, K&$\bar \omega /\theta
 _D$\\
\hline
 Al&3&12.2&6.4&1.25&1.48&26&1.97&382&394&0.97\\
 Zn&2&22.6&4.17&0.51&0.85&30&1.75&264&234&1.13\\
 Pb&4&89.4&2.16&1.11&1.97&10&15.55&28&94.5&0.30\\
 Sn&4&50.9&3.32&1.12&1.29&45&6.25&48&170&0.28\\
\end{tabular}
\end{ruledtabular}
\end{table*}

\begin{figure}
\includegraphics{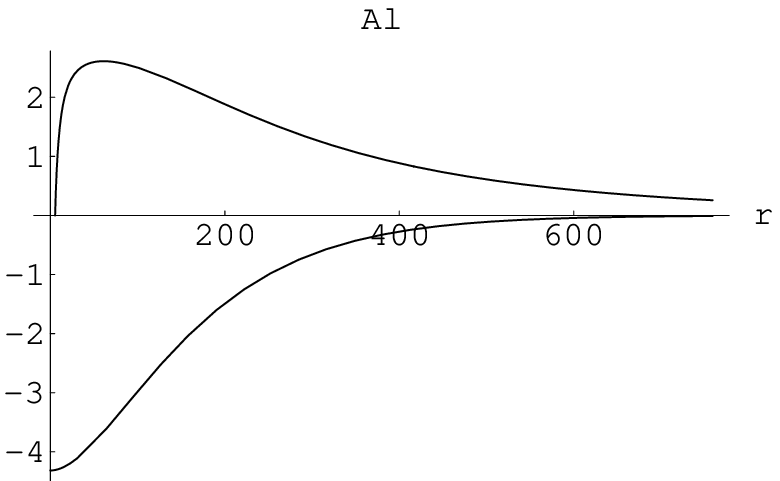}
\caption{\label{fig:epsart} Calculated potential $V\left( r \right)$
(lower semiplane) and WF $\psi \left( r \right)$ (upper semiplane)
for bound state in Al. Values of $V$ are given in degrees Kelvin,
$r$ is in Angstroms and $\psi$ in arbitrary units.}
\end{figure}
\begin{figure}
\includegraphics{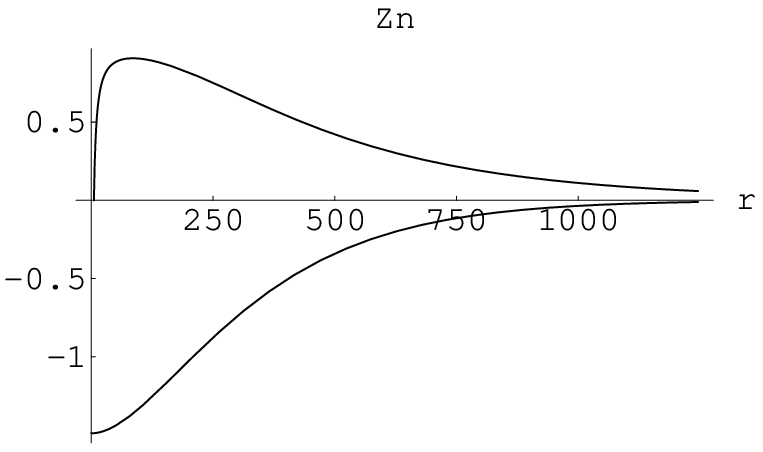}
\caption{\label{fig:epsart} Calculated potential $V\left( r \right)$
(lower semiplane) and WF $\psi \left( r \right)$ (upper semiplane)
for bound state in Zn. Values of $V$ are given in degrees Kelvin,
$r$ is in Angstroms and $\psi$ in arbitrary units.}
\end{figure}
\begin{figure}
\includegraphics{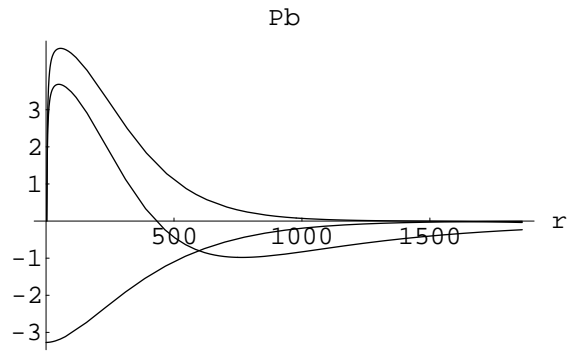}
\caption{\label{fig:epsart} Calculated potential $V\left( r \right)$
(lower semiplane) and WFs $\psi \left( r \right)$ (left-hand sides
of the curves are in the upper semiplane) for two bound states in
Pb. Values of $V$ are given in degrees Kelvin, $r$ is in Angstroms
and $\psi$ in arbitrary units.}
\end{figure}
\begin{figure}
\includegraphics{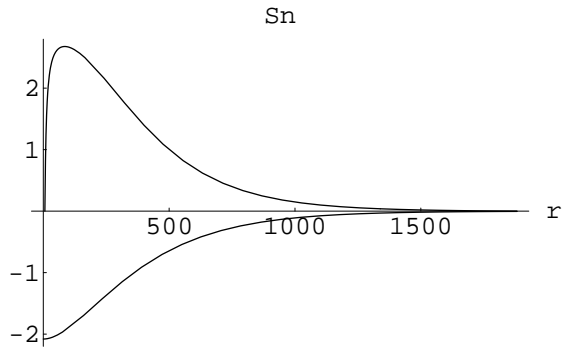}
\caption{\label{fig:epsart} Calculated potential $V\left( r \right)$
(lower semiplane) and WF $\psi \left( r \right)$ (upper semiplane)
for bound state in Sn. Values of $V$ are given in degrees Kelvin,
$r$ is in Angstroms and $\psi$ in arbitrary units.}
\end{figure}

\section{CONCLUSIONS}

The following circumstance attracts attention. The data on the
superconductivity gap for the considered metals contradict the
natural consequences of the BCS theory. Indeed, according to this
theory, the higher is the value of $\theta _D$ the larger should be
$\Delta _0$; now comparing, \emph{e.g.}, the Table 1 data for Al and
Zn with those for Pb and Sn does not confirm this prediction. In
this situation, a study of beryllium, for which $\theta _D  = 1000$
K, and the transition temperature $T_c = 0.03$ K, would be of an
incontestable interest.

In the framework of the BCS theory, such deviations from its
predictions are compensated by considering the value of the
so-called Coulomb pseudopotential $\mu ^*$, which is used as a
fitting parameter \cite{10}. This is due to impossibility of a
satisfactorily precise evaluation of $\mu ^*$ without a comparison
with the data on transition temperature or superconductivity gap.

On the other hand, as specified above, the value of $\bar \omega$
can be estimated within the framework of the new theory after a
correct processing of the CVC of TJ; therefore, a final version of
superconductivity gap calculation for PT metals can be totally free
of fitting parameters. From this point of view, the proposed theory
appears to be more advantageous than the BCS theory.

We may note as a conclusion that a calculation of the gap value for
superconductors produced from semiconductors can be performed
without determining the parameter $\bar \omega$ at all. Indeed,
assuming that the frequency of electron transitions induced by a
scattering on an ionized impurity is substantially higher that the
frequency of EPI-caused spontaneous transitions, the value of $
\delta p$ can be estimated, \emph{e.g.}, based on Hall mobility
data. In this case, considering the superconductors obtained from
semiconductors could provide the most convincing evidence in favor
of our theory.

\begin{acknowledgements}
The author expresses his gratitude to Prof. A. A. Rukhadze  for his
unfailing attention towards the problem and a moral support.
\end{acknowledgements}

\appendix
\section{Calculation of the Effective EEI Potential in the Case of Coulomb Interaction}

Self-consistency equations for the effective EEI potential
simplified based on the approximate evaluation of integrals within
Fermi sphere \cite{6} for this case can be expressed in the form of:
\begin{eqnarray}
\begin{array}{l}
 \tilde U_{\bf q}^{{\bf p},{\bf k}}  = \frac{{2\pi e^2 }}{{\varepsilon q^2 }} \\
  - \frac{{e^2 m}}{{\varepsilon \pi ^2 q^3 }}I\left( {\frac{{{\bf pq}}}{q}} \right)\tilde U_{\bf q}^{0,{\bf p} + {\bf q}}  + \frac{{e^2 m}}{{\varepsilon \pi ^2 q^3 }}I\left( {\frac{{{\bf pq}}}{q} + q} \right)\tilde U_{\bf q}^{{\bf p},0}  \\
  - \frac{{e^2 m}}{{\varepsilon \pi ^2 q^3 }}I\left( {\frac{{{\bf kq}}}{q} - q} \right)\tilde U_{\bf q}^{0,{\bf k}}  + \frac{{e^2 m}}{{\varepsilon \pi ^2 q^3 }}I\left( {\frac{{{\bf kq}}}{q}} \right)\tilde U_{\bf q}^{{\bf k} - {\bf q},0}  \\
  + \frac{{e^2 m}}{{2\varepsilon \pi ^2 q^3 }}I\left( {\frac{{{\bf pq}}}{q}} \right)\tilde U_{\bf p}^{0,{\bf p} + {\bf q}}  - \frac{{e^2 m}}{{2\varepsilon \pi ^2 q^3 }}I\left( {\frac{{{\bf pq}}}{q} + q} \right)\tilde U_{{\bf p} + {\bf q}}^{0,{\bf p}}  \\
  - \frac{{e^2 m}}{{2\varepsilon \pi ^2 q^3 }}I\left( {\frac{{{\bf kq}}}{q}} \right)\tilde U_{\bf k}^{0,{\bf k} - {\bf q}}  + \frac{{e^2 m}}{{2\varepsilon \pi ^2 q^3 }}I\left( {\frac{{{\bf kq}}}{q} - q} \right)\tilde U_{{\bf k} - {\bf q}}^{0,{\bf k}} , \\
 \end{array}
\end{eqnarray}
where $I\left( x \right) = \pi \left( {K_F^2  - x^2 } \right)\ln
\left| {\frac{{K_F  - x}}{{K_F  + x}}} \right| - 2\pi K_F x$ and the
following relationships are satisfied:
\begin{equation}
\begin{array}{l}
 \tilde U_{\bf q}^{{\bf p},{\bf k}}  = \tilde U_{ - {\bf q}}^{{\bf k},{\bf p}} , \\
 \tilde U_{\bf q}^{{\bf p},{\bf k}}  = \tilde U_{ - {\bf q}}^{{\bf p} + {\bf q},{\bf k} - {\bf q}} , \\
 \tilde U_{\bf q}^{{\bf p},{\bf k}}  = \tilde U_{ - {\bf q}}^{ - {\bf p}, - {\bf k}} .
 \end{array}
\end{equation}
It follows immediately that:
\begin{equation}
\tilde U_{\bf q}^{0,0}  = \frac{{2\pi e^2 }}{{\varepsilon q^2  +
\lambda ^2 }},
\end{equation}
where $\lambda ^2  = 4e^2 mK_F/\pi $.

Henceforth, we restrict ourselves to the case $p,k \approx K_F$,
\emph{i.e.} that of electron interaction in the vicinity of Fermi
surface. Then, considering that $\varepsilon K_F^2 \gg \lambda ^2$,
the equation for $\tilde U_{\bf q}^{{\bf p},0}$ can be simplified
omitting a series of terms $ \sim {{e^2 }\over{\varepsilon K_F^2}}$.
As a result, it becomes as follows:
\begin{eqnarray}
\left[ {1 - \frac{{me^2 }}{{\varepsilon \pi ^2 q^2 }}I'\left(
{\frac{{{\bf pq}}}{q} + \frac{q}{2}} \right)} \right]\tilde U_{\bf
q}^{{\bf p},0}= \frac{{2\pi e^2 }}{{\varepsilon q^2 }} + \frac{{me^2
}}{{2\varepsilon \pi ^2 q^3 }}I\left( q \right)\tilde U_{\bf
q}^{0,0} .
\end{eqnarray}
Now take into account that the value $ \left\langle {I'\left(
{{{{\bf pq}} \over q} + {q \over 2}} \right)} \right\rangle $
averaged over the angles between the vectors ${\bf p}$ and ${\bf q}$
is approximately equal to $- 4\pi K_F$ at $q \to 0$. Then, simple
calculation yields:
\begin{equation}
\tilde U_{\bf q}^{{\bf p},{\bf k}}  = 2\pi e^2 \frac{{\varepsilon
q^2 }}{{\left( {\varepsilon q^2  + \lambda ^2 } \right)^2 }},
\end{equation}
which, on the whole, agrees with the present-day concept of the EEI
effective potential, as $\tilde U_{\bf q}^{{\bf p},{\bf k}}  \sim
e^2 /\lambda ^2$ at $q \sim \lambda /\sqrt \varepsilon$.

\end{document}